\documentclass[%
reprint,
 amsmath,amssymb,
pre,
floatfix,
]{revtex4-2}
\usepackage{graphicx}
\usepackage{dcolumn}
\usepackage{bm}
\usepackage{hyperref}
\usepackage{siunitx}
\usepackage{layouts}
\usepackage{amsmath}  
\usepackage{amssymb,amsthm}
\usepackage[dvipsnames]{xcolor}
\usepackage{ulem}
\usepackage{natbib}
\usepackage{mathrsfs}
\usepackage{subfigure}
\usepackage{picture}
\usepackage{color}
\usepackage{mathrsfs}
\usepackage{ulem}

\definecolor{cbc1}{HTML}{332288}
\definecolor{cbc2}{HTML}{117733}
\definecolor{cbc3}{HTML}{44AA99}
\definecolor{cbc4}{HTML}{88CCEE}
\definecolor{cbc5}{HTML}{DDCC77}
\definecolor{cbc6}{HTML}{CC6677}
\definecolor{cbc7}{HTML}{AA4499}
\definecolor{cbc8}{HTML}{882255}
\definecolor{gnup_blue3}{HTML}{57B5E8}
\definecolor{gnup_purple1}{HTML}{9400D4}
\definecolor{gnup_orange}{HTML}{FFA600}

\definecolor{dartmouthgreen}{rgb}{0.05, 0.5, 0.06}
\def\Pe{\mbox{\rm Pe}}

\def\bbox#1{\ifmmode \mbox{\boldmath $#1$} \else \mbox{${\boldmath #1}$} \fi}

\graphicspath{../figures/}
\usepackage[T1]{fontenc}

\begin{document}

\preprint{APS/123-QED}

\title{Residence time distributions for in-line chaotic mixers}

\author{Nelson Pouma\"ere}
\author{Beno\^it Pier}
\author{Florence Raynal}
\affiliation{Laboratoire de Mécanique des Fluides et d'Acoustique, Univ Lyon, \'Ecole centrale de Lyon, INSA Lyon, Univ Claude Bernard Lyon 1, CNRS, F-69134 \'Ecully, France}

\date{\today}

\begin{abstract}
We investigate the distributions of residence time for in-line chaotic mixers; in particular, we consider the Kenics\textsuperscript{\textregistered}, the F-mixer and the Multi-level laminating mixer (MLLM), and also a synthetic model that mimics their behaviour and allows exact mathematical calculations. 
We show that whatever the number of elements of mixer involved, the distribution possesses a $t^{-3}$ tail, so that its shape is always far from Gaussian. 
This $t^{-3}$ tail also invalidates the use of second-order moment and variance. 
As a measure for the width of the distribution, we consider the mean
absolute deviation and show that, unlike the standard deviation, it converges in the limit of large sample size.
Finally, we analyse the performances of the different in-line mixers from the residence-time point of view when varying the number of elements and the shape of the cross-section. 
\end{abstract}

\maketitle

\section{Introduction}

Efficient stirring is the key ingredient of good mixing.
This mechanism is generally associated with a turbulent flow, but even when the flow-field is laminar, dynamical systems theory allows chaotic trajectories by stretching and folding of fluid elements, a process called chaotic advection \cite{bib:ottino1990,bib:Aref_etal_RevModPhys2017,bib:gouillart_etal2011}. 
Chaotic advection arises in a large diversity of natural or industrial flows.
Extreme examples are mixing in geophysical flows (in the oceans \cite{bib:budyansky_etal2009}, or magma in the earth mantle \cite{bib:rossi2017}), where the typical length scale reaches hundreds of kilometers, and microfluidics \cite{bib:wigginsottino04,bib:bruus2008}, with typical length scale of the order of \SI{100}{\micro\metre}, that is, 9 orders of magnitude smaller.   



In this article, we are interested in in-line mixers, consisting of a succession of identical elements, which have applications from millifluidics \cite{bib:Creyssels_etal2015,bib:bahrani2019} to microfluidics \cite{bib:stroocketal02}. 
Although solving the concentration field is not easy to achieve because of their complicated geometry \cite{bib:gorodetskyi_etal2014,bib:borgogna2019},   
it is well known that those mixers achieve a very good mixing by reproducing the baker's map. 
Thus they can indeed be considered as ideal mixers.

The present investigation focuses on another aspect of in-line mixers, their residence-time distributions or RTD \cite{bib:danckwerts1953,bib:danckwerts1958}: 
an ideal mixer  is characterized by a very narrow Gaussian or a Dirac centered on the mean travel time.
However, when considering only one element of an in-line mixer, the histogram of residence time is very broad and often monotonously decaying, with a maximum equal (or very close) to the minimum time involved to cross the element \cite{bib:Raynal_Carriere_2015}: a behaviour very far from that of an ideal mixer. 
Our goal is thus to study how the histogram evolves when increasing the number of elements. 

Residence time distribution is a complex feature, not always correctly comprehended. 
Indeed, let us consider the case of the flow in a cylindrical pipe with circular cross-section. 
The parallel flow-field in the $x$ direction is a parabolic profile of equation:
\begin{equation}
    v_x(r)=2\,v_m\,(1-r^2/R^2)\, 
    \label{eq:velocity_Poiseuille_circular}
\end{equation}
where $r$ is the radial distance to the center of the section, $R$ is the radius of the pipe, and $v_m$ the mean velocity over the section. 
Because of the cylindrical symmetry, the residence time $t$ depends only on $r$ as
\begin{equation}
    t(r)=L/v_x(r),
    \label{eq:time_V}
\end{equation}
for a section of length~$L$.
Suppose now that we calculate the mean residence time $t_m$ just by sampling randomly $M$ particles at the inlet section at $t=0$ (what Danckwerts named a ``pulse signal'' \cite{bib:danckwerts1958}), and measure the mean of the $M$ corresponding residence times $t$. 
The result should be the same as what is obtained from the continuous equation:
\begin{eqnarray}
    t_m&=&\frac{1}{\pi R^2}\int_0^R t(r)\, 2\pi r dr\\
    &=& \frac{L}{2v_m}\int_0^R \frac{1}{1-r^2/R^2}\, \frac{2 r dr}{R^2}\\
    &=& \frac{L}{2v_m}\int_0^1 \frac{1}{1-u}\,du\,,
\end{eqnarray}
where we have used equation (\ref{eq:time_V}) and set $u=r/R$. 
Finally $t_m$ diverges logarithmically when $u$ approaches $1$ ($r$ approaches $R$), so that the mean time calculated this way is not defined.  
The reason lies in the way the mean time is calculated: 
when considering the inlet section during a lapse of time $dt$, many more particles cross at the center (where the velocity is maximal) than near the walls (where the velocity is very weak). 
As expressed by Danckwerts \cite{bib:danckwerts1953}, \textit{``there is a variation in velocity from the axis to the wall of the pipe, so that the central "core" of fluid moves with a velocity greater than the mean, while the fluid near the wall lags behind.''}
In order to calculate a mean time,  this non-uniform flux of particles must be taken into account, by properly weighting the statistics \cite{bib:raynal_etal2013,bib:Raynal_Carriere_2015, bib:oteski_Duguet_Pastur2014}. 
As the quantity of particles that cross a section during $dt$ is proportional to the crossing velocity, the weight must also be chosen proportional to this velocity, \textit{i.e.} $v_\perp/v_m$ where $v_\perp$ is the component of the velocity perpendicular to the cross section. 
Now calculating again the mean time $t_m$ using this weight, with $v_\perp=v_x$, leads to the trivial expression
\begin{equation}
    t_m=\frac{1}{\pi R^2}\int_0^R \frac{v_x(r)}{v_m}\,t\, 2\pi r dr=L/v_m
    \label{eq:tm_L_vm}
\end{equation}
because of equation \ref{eq:time_V}. 
We finally obtain the desired result 
\begin{equation}
    t_m=\frac{\cal V}{Q}
\end{equation}
where ${\cal V}$ is the volume of an element and  $Q$ the flow-rate. 

In a former article \cite{bib:Raynal_Carriere_2015}, we proposed to use the time of flight in order to obtain statistics of residence time. 
The time of flight is the lapse of time between the inlet and outlet of a given element when following a single fluid particle. 
Unlike RTD, the time of flight is a Lagrangian quantity, very close to the time of first return \cite{bib:eckmann_ruelle1985}, or to the waiting time (time spent by a particle in a given domain ${\cal D}$) \cite{bib:artuso_etal2008}, both introduced for dynamical systems. 
 Obviously, a particle trajectory is more likely to enter a given element in regions of high velocity than near the walls, so that there is no need for weighting the statistics as for RTD: 
when averaged, the time of flight converges naturally toward the mean time $t_m={\cal V}/Q$ \cite{bib:Raynal_Carriere_2015}. 

In the following, we will use time of flight to construct residence time distributions. 
The flow field is laminar, and we mostly consider non diffusive particles, which corresponds to flows at high P\'eclet numbers on short times, for which the effects of molecular diffusion are negligible. 
The mean residence time in $n$ elements is denoted by $t_m^{(n)}=n\times t_m$, where $t_m\equiv t_m^{(1)}$ is the mean residence time in a single element of mixer; 
similarly, $t_{min}^{(n)}=n\times t_{min}$ denotes the minimal time taken by a particle to cross $n$ elements;  
the maximum time is infinite, due to the zero-velocity field on the walls. 
The density probability of residence time in $n$ elements is denoted by $f_n(t)$. 

This paper is organized as follows: in the next section we present the different mixers studied. 
We begin with the real mixers, and show that their auto-correlation coefficient decreases very rapidly with the number of elements. 
This allows us to introduce a kinematic model that mimics the residence time distributions in a single element. 
In the following section we vary the number of mixing elements from $1$ to $n$. 
In particular, we show that $t^{-3}$ tail that exists for $1$ element persists when the number of elements is increased. 
Then we explain how, because of this $t^{-3}$ tail, the use of the classic standard deviation is forbidden. 
We thus discuss how to measure the stretching of RTD, and choose the mean absolute deviation;  we can therefore compare the different in-line mixers. 
Finally in the last section we use this tool to discuss the influence of the cross-section geometry of mixing elements in the stretching of RTD.\\ 


\section{Mixers studied}
The mixers studied here ---the Kenics\textsuperscript{\textregistered}, the F-mixer and the Multi-level laminating mixer--- enable global chaos \cite{bib:Raynal_Carriere_2015};
they are constituted of $n$ identical elements. 
For each mixer we calculate the RTD using time of flight: we follow a fluid particle over time, and record the time taken to cross each element. 
For the calculation of the time of flight in $n$ elements, we sum the $n$ individual times of flight corresponding to $n$ elements in a row.

\vskip2mm
\onecolumngrid

\unitlength=1mm
\begin{figure}[hb!]
    \centering
\includegraphics{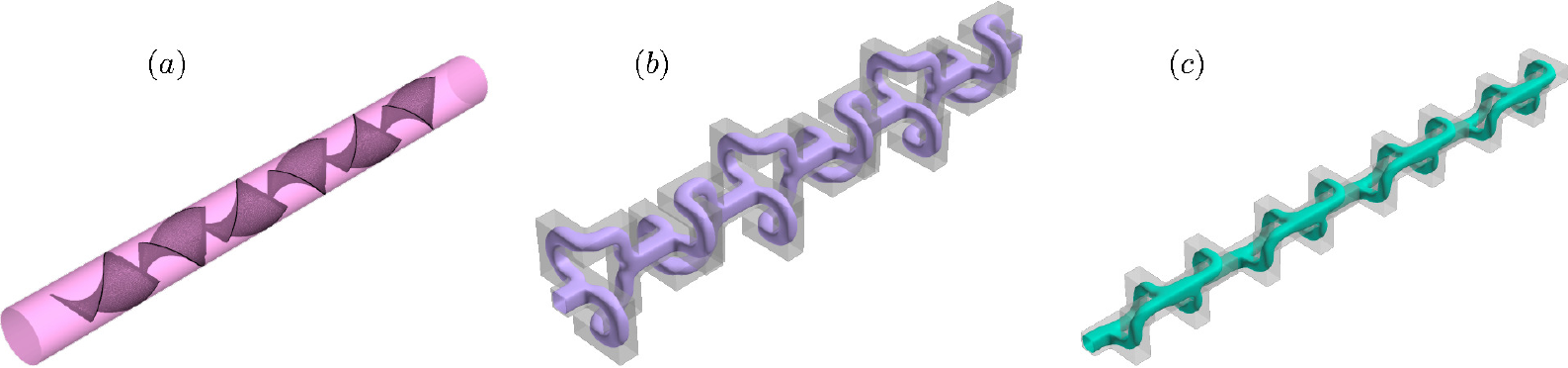}
    \caption{Computational geometry of the different mixers studied: $(a)$ the Kenics$^{\textrm{\textregistered}}$ mixer (6 elements); $(b)$ the Multi-level laminating mixer (6 elements); 
$(c)$ the ”F” mixer (8 elements). For $(b)$ and $(c)$ an iso-surface of velocity modulus for a Stokes flow is plotted in color.}
    \label{fig:real_mixer_drawing}
\end{figure}

\bigskip
\hrule
\bigskip

\twocolumngrid

\subsection{Real mixers}
The numerical treatment of the velocity field by finite element method, and integration of the trajectories by a fourth-order Runge-Kutta method for the mixers studied here was explained in detail in \cite{bib:Raynal_Carriere_2015}; 
we use the same numerical data here.


The computational geometries for three mixers are depicted in figure \ref{fig:real_mixer_drawing}. 
The corresponding Poincar\'e sections and Lyapunov exponents are not shown here, but can be found in \cite{bib:Raynal_Carriere_2015}. 
A particle which exits at the outlet cross-section of a computational geometry is reintroduced at the same location in the inlet cross-section. 
This enables to follow a particle on a very long number of elements, and we note the consecutive residence time in each element. 
Note that the number of elements involved in the computational geometries is not significant in this study. 

For each mixer 4 long trajectories were calculated.
A trajectory is terminated when the point ends in a wall, which may happen due to  intrinsically limited numerical accuracy, or when a point is so close to a wall that the time taken to escape the element is too high. 
For this work the loss of particles is less than 1\% \cite{bib:Raynal_Carriere_2015}

The Kenics$^{\textrm{\textregistered}}$ mixer 
\cite{bib:hobbsetal98} is composed of a series of identical internal blades inside a circular pipe; 
each blade has a helical shape, alternately right- or left-handed, and the leading edge of a given blade is at right angle of the trailing edge of the preceding blade. 
The computational geometry used here is shown in figure \ref{fig:real_mixer_drawing}$a$: note that six elements are represented, so that the periodicity of the flow arises after 2 elements. 

The multi-level laminating mixer (MLLM) \cite{bib:grayetal99,bib:carriere2007,bib:anxionnaz2017} has a three-dimensional configuration intended to mimic the baker's map. 
The computational geometry used is shown in figure \ref{fig:real_mixer_drawing}$b$, with six elements represented. 
The successive elements are inverted so as to break the symmetry of the flow and avoid small residual non chaotic regions \cite{bib:carriere2007}. 
Therefore here again, the structure has a periodicity of two elements. 

Finally, the F-mixer \cite{bib:chenmeiners04,bib:chenetal09} has a similar topological behaviour as the MLLM, although its geometry is simpler; compared to the former, it is less symmetric, which is not a problem for Stokes flows. 
Indeed, its Lyapunov exponent is, as for the MLLM, equal to $\ln2$ \cite{bib:Raynal_Carriere_2015}. 
Its computational geometry is represented in figure \ref{fig:real_mixer_drawing}$c$, with eight elements. 
However, compared to the former, a mixing element represents a whole spatial period of the mixer. 
This property will be taken into account later.



\subsection{Auto-correlation coefficient}
How is a time of flight of a given element correlated to the time of flight in an element further away? 
It can be estimated through the auto-correlation coefficient,
\begin{equation}
  R(i) = \frac{M}{M-i}\, \frac{\sum_{j=1}^{M-i} (t_j - t_m)(t_{j+i} - t_m)}{\sum_{j=1}^{M} (t_j - t_m)^2}\,;   
\end{equation}
here $i=1$ corresponds to the correlation between two consecutive elements. 
\begin{figure}
\includegraphics{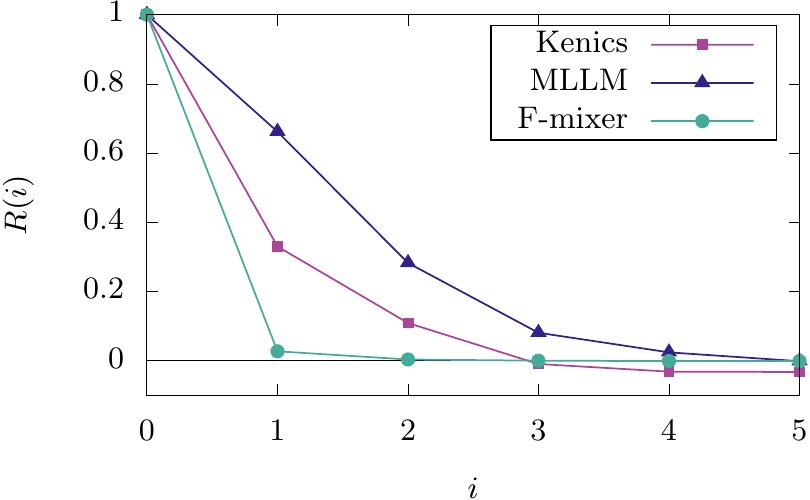}
    \caption{Evolution of the correlation coefficient $R(i)$ between residence time values in elements that are $i$ elements away, for the three real mixers.}
    \label{fig:correlations}
\end{figure}
In figure \ref{fig:correlations} we have plotted the auto-correlation coefficient for the three mixers depicted above. 
As can be seen the time of flight decorrelates very rapidly with the number of elements. 

The decorrelation is the fastest for the F-mixer. 
Indeed, unlike the MLLM, its asymmetry leads to very different times of flight depending on the branch chosen in an element. 
Furthermore, as already noted, one element of the F-mixer corresponds to a full spatial period, in contrast to the two other mixers. 
But, even when considering this particularity, the decorrelation is still the fastest, since $R(1)$ is nearly zero, thus below $R(2)$ for the two other mixers. 

Overall, for all mixers, the time of flight is totally decorrelated after only four basic elements. 
This rapid decorrelation of time of flight justifies a priori the model that we present hereafter.

\subsection{A residence time model}
\label{sec:residence_time_model}
We propose to model residence time in such mixers using the time of flight between inlet and outlet of an element with simple geometry. 
Such a model was previously used to model the distribution of time of flight in a single element of mixer \cite{bib:Raynal_Carriere_2015}. 
It can be described as follows:
\begin{enumerate} 
\item the flow through one element of the mixer is modeled by a non-chaotic flow possessing no-slip boundaries (for instance a piece of pipe with circular cross-section);  
\item the effect of global chaos on the trajectory of the fluid particle is modeled by random reinjection at the entry to the next element with a probability density taking into account the fact that the particle randomly samples the whole section, but less near the walls; 
\item in order to conserve mass, as explained in the introduction, the probability density function of the location of reinjection is taken proportional to the local velocity (see Eq. \eqref{eq:f1dt} below for a pipe with circular cross-section).
\end{enumerate}

In the following, we mostly focus on the case of a circular cross-section (other shapes are also considered, see section \ref{sec:discussion}). 
In practice we generate random numbers with a parabolic probability density using an inversion method \cite{bib:devroye1986}, see appendix  \ref{app:inversion_method}. 

The circular cross-section enables indeed an analytical expression for the probability density $f_1(t)$ to have a time of flight of duration $t$ for 1 element: 
the probability to have a duration of time in between $t$ and $t + dt$ is equal to that of having a particle reinjected in between $r$ and $r + dr$, where $t$ and $r$ are linked by relation (\ref{eq:time_V}):
\begin{equation}
    f_1(t)\, dt= \frac{v_x(r)}{v_m}\,\frac{2\pi r\, dr}{\pi R^2}\,,
    \label{eq:f1dt}
\end{equation}
where $v_x(r)$ verifies equation (\ref{eq:velocity_Poiseuille_circular}). 
When differentiating equation (\ref{eq:time_V}), we obtain
\begin{equation}
    -\frac{2r\,dr}{R^2}=-\frac{L}{2v_m\,t^2}dt
\end{equation}
which, when combined with equations (\ref{eq:f1dt}), (\ref{eq:time_V}) and (\ref{eq:tm_L_vm}) leads to
\begin{equation}
    f_1(t)=\frac{t_m^2}{2t^3}
    \label{eq:f1_Poiseuille_circular}\,.
\end{equation}
This is indeed the profile  obtained numerically for 1 element, see figure \ref{fig:pdf_model}\,$a$. 
Not surprisingly, the expression derived by Danckwerts \cite{bib:danckwerts1953} is recovered. 
This $t^{-3}$ tail was also found for the three mixers in the case of a single element ($n=1$).
Because large times of flight correspond to points located near the wall where the velocity is weak, this behaviour was related to the region of constant shear near the wall \cite{bib:Raynal_Carriere_2015}. 
An indirect proof can be found when considering the plane Couette flow, where the shear is constant everywhere: 
for this flow also, the probability density follows equation (\ref{eq:f1_Poiseuille_circular}) \cite{bib:Raynal_Carriere_2015}. 

In the following, we propose to use this model for $n$ consecutive elements of an in-line mixer. 

\section{Residence time distributions: from 1 to \texorpdfstring{$n$\ }\ mixing elements}
For a single-element of mixer, the RTD is characterized by the following properties \cite{bib:Raynal_Carriere_2015}:\\
 - The existence of a $t^{-3}$ tail;\\
 - A maximum close to $t=t_{min}$. \\
As already stated, our idea is now to go further and explore the more realistic case of multiple elements. 

\subsection{Model}
The model is of particular interest since, because of its intrinsic simplicity, it allows to increase arbitrarily the number of elements. 
\begin{figure}
\includegraphics{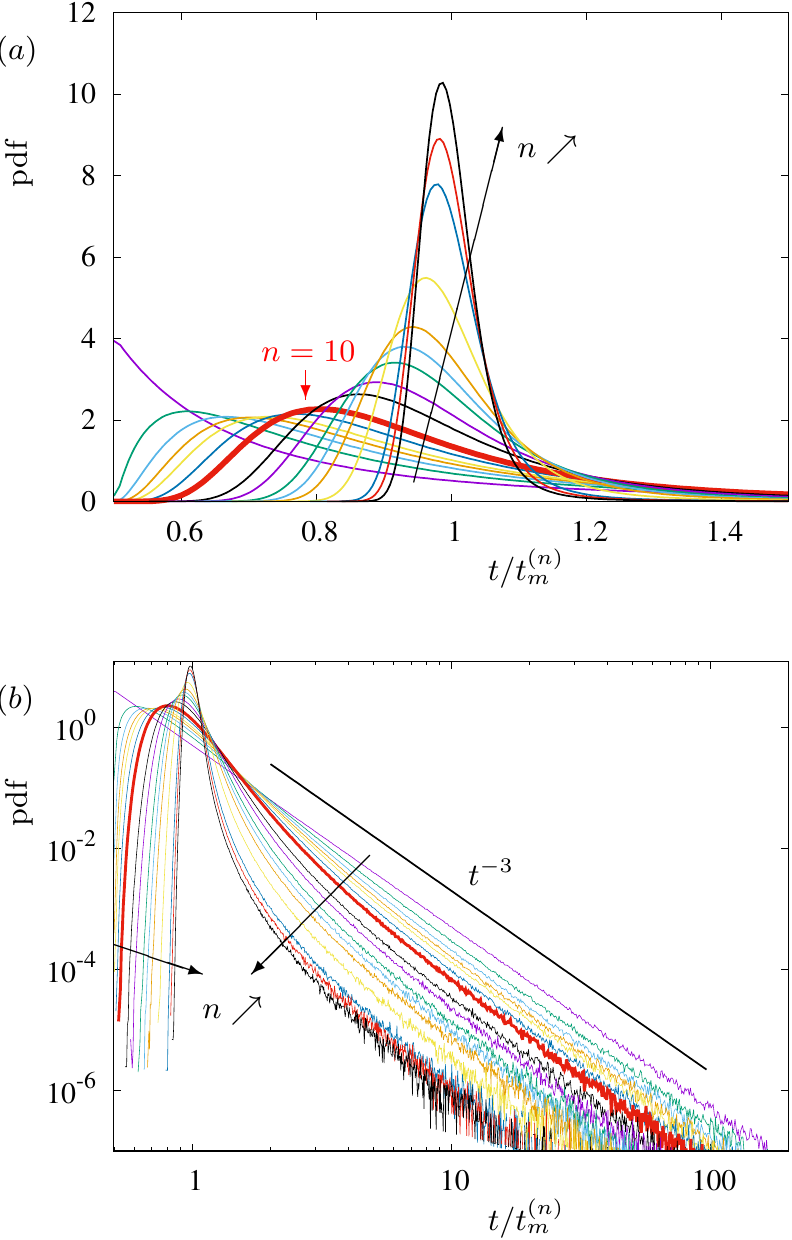}
\caption{Residence time distribution $f_n$ for the model chaotic flow (Poiseuille flow with circular cross-section), for different numbers of sections, $n=1$, 2, 3, 4, 5, 7, 10, 14, 20, 30, 50, 70, 100, 140, 200, 300, 500, 700, and 1000. 
By way of comparison with real mixers, the case $n=10$ is drawn thicker. 
Each distribution was built with $M=10^8$ data; 
a data is the sum of $n$ independent residence times. 
$(a)$: linear scale; $(b)$: logarithmic scale. 
Even for a very large number of sections $n$, the pdf is far from Gaussian, and exhibits a $t^{-3}$ power-law tail. 
For the large number of elements, the tails are slightly more scattered, because more points are in the peak (figure \ref{fig:pdf_model}$a$).
\label{fig:pdf_model}
}
\end{figure}
In figure \ref{fig:pdf_model}$a$ we show the non-dimensional time distributions (built as a non-dimensional pdf) for a number of elements varying from $n=1$ to $n=1000$.
Of course $1000$ elements is not a realistic configuration in practice, but it allows to visualize theoretically the rate of convergence towards the ``perfect'' mixer. 

The first notable point is that for $n\ge 2$ the distribution is actually a bell curve, with a maximum different from $t=t_{min}$, therefore a much improved shape compared to the $n=1$ case. 
When $n$ increases the curve becomes more peaked, and the position of the maximum tends to the mean time of flight $t_m^{(n)}=n\times t_n^{(1)}$. 
However the convergence is very slow.  
The case $n=10$, that can be considered as a reasonable maximum number of elements in a real mixer, is shown as a thicker line (in red): 
as can be seen, the distribution is still very broad; 
furthermore, even for $n=1000$, the maximum of the distribution is still not completely centered on the mean time. 

The second notable point is visible in the log-log plot of the same distributions (figure \ref{fig:pdf_model}$b$):
the $t^{-3}$ tail that was found for $n=1$ persists at all higher values of $n$,  and the distributions remain very asymmetric. 
In the model, all residence times in an element are completely independent of each other. 
It can be shown that the distribution of the sum of two decorrelated data with an algebraic tail also possesses an algebraic tail \cite{bib:feller1971,bib:wilke_etal1998}. 
In appendix \ref{app:tail}, we apply this result and prove the existence of this $t^{-3}$ tail when summing $n$ independent data taken from the same distribution with a $t^{-3}$ tail. 

In real mixers, two consecutive times are not completely decorrelated as in the model (figure \ref{fig:correlations}). 
However, because the correlation is weak, quite similar results are expected. 


\subsection{Mixers}
\begin{figure}[h]
\includegraphics[]{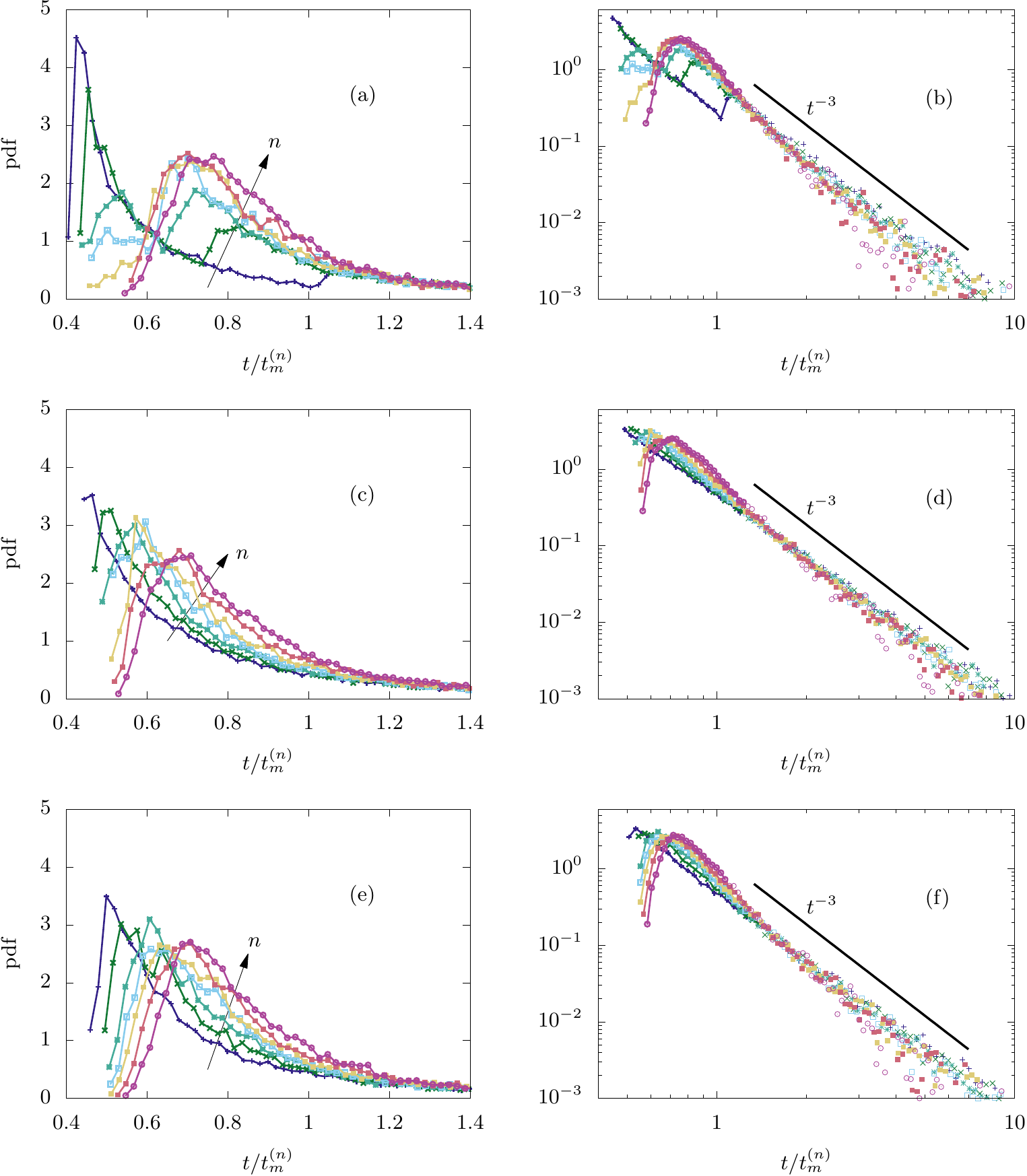}
\caption{Residence time distributions for the three real mixers for different numbers of elements, $n=$ 1, 2, 3, 4, 5, 7, 10. From top to bottom: Kenics$^{\hbox{\textregistered}}$, MLLM, F-mixer. 
For each mixer, 4 trajectories were calculated, corresponding to a total of 16886 times of flight for the Kenics$^{\hbox{\textregistered}}$, 33570 for the MLLM, and 18987 for the F-mixer. 
We used a sliding average, so that the number of points are roughly the same for the different values of $n$.
Left: linear scale; Right: logarithmic scale. 
As for the model, the tail is more noisy for the highest values of $n$ ($n\ge7$): the weight (integral under the curve) of the bell-shaped part is more significant, which implies that the proportion of points in the tail is less important. }
\label{fig:RTD_real_mixers}
\end{figure}

Figure \ref{fig:RTD_real_mixers} shows the RTD for the three mixers.
Due to a much reduced number of data points for the real mixers compared to the model, the histograms are limited to distributions for $n=10$ elements;
anyhow, most in-line mixers have less than $10$ mixing elements.

As expected, the distributions are quite similar to what was obtained with the model, although not as smooth, due to the much smaller sample of data. 
As for the model, the distributions are still broad for $n=10$, and quite far from the desired Gaussian shape. 
Another important point is the persistence of the $t^{-3}$ tail, visible on the log-log plot. 
This is not surprising: we demonstrated that summing $n$ independent variables with a $t^{-3}$ tail led to a distribution with a similar tail. 
These real-mixer data are poorly correlated (see figure \ref{fig:correlations}), so that the variables may be considered as nearly independent. The assumption of uncorrelated data is almost exact for the F-mixer, for which the auto-correlation coefficient has fallen to negligible values after only one element. 
Moreover the least noisy tail is that of the MLLM (figure \ref{fig:RTD_real_mixers}$d$), for which we have twice as much data as for the two other mixers, but that also corresponds to the more correlated mixer. 
Finally, note that El Omary \textit{et al.} \cite{bib:el_Omary_Younes_Castelain2021} also found a $t^{-3}$ tail when properly weighting their statistics.



A distribution with an algebraic tail $t^{-\alpha}$ 
(also called Pareto distribution) belongs to the family of ``heavy-tailed'' distributions \cite{bib:klebanov2003}. 
This type of distribution is well known in economy \cite{bib:pareto1964}, finance \cite{bib:rachev2003}, physics \cite{bib:srokowski2009}, maths \cite{bib:ramsay2006} and even bibliometry~\cite{bib:lariviere2016}. 

\onecolumngrid

\bigskip\smallskip
\hrule
\bigskip
\twocolumngrid

\section{A measurement tool for the stretching of residence time distribution}

\subsection{Why not use the standard deviation?}
When dealing with distributions it is natural to measure the histogram width. 
Because many distributions in fluid mechanics are Gaussian, or close to Gaussian, it is usual to use the standard deviation, or even higher moments. 
In our case, the standard deviation for $n$ consecutive elements is denoted by $\sigma_2^{(n)}$ and defined as:
\begin{equation}
 \sigma_2^{(n)}=\sqrt{\int_{t_\mathrm{min}^{(n)}}^\infty f_n(t)\,\left(\frac{t}{t_m^{(n)}}-1\right)^{2} dt}\,.  
\end{equation}
However, because of the $t^{-3}$ tail, the integral diverges and this quantity is clearly not well-defined.

It is always possible in practice to calculate a standard deviation from a series of $M$ values of time of flight as:
\begin{equation}
 \sigma_2^{(n)}=\sqrt{\frac{1}{M}\,\sum_{j=1}^M \left(\frac{t_j}{t_m^{(n)}}-1\right)^{2}}\,.
\end{equation}
Note again that, because we deal with times of flight (resulting from a single trajectory), the weighting is here naturally included in the statistics. 

\begin{figure}
    \includegraphics[]{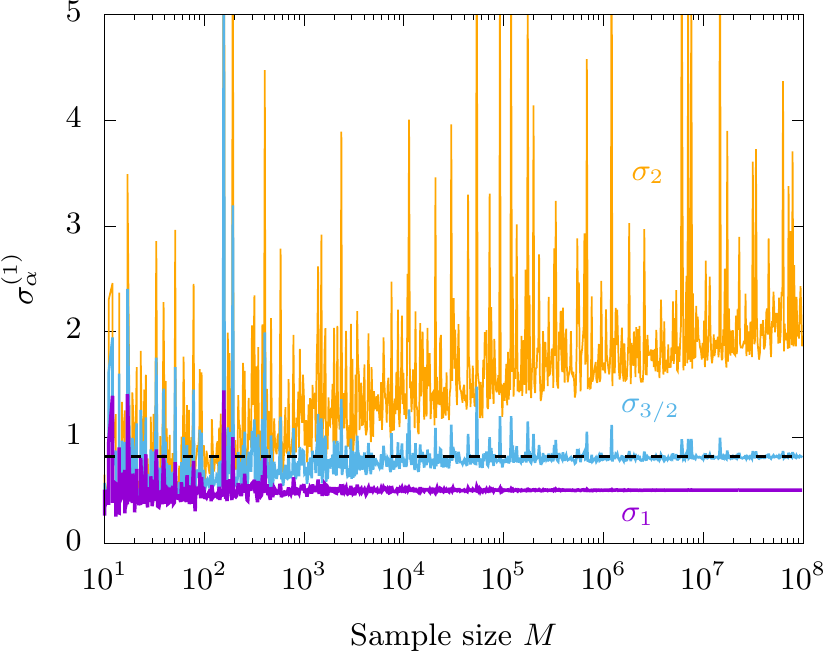}
    \caption{Evolution of the standard deviation $\sigma_2$ (orange), centered absolute moment of order 3/2 $\sigma_{3/2}$ (light blue), and mean absolute deviation $\sigma_{1}$ (purple) for one element of the Poiseuille flow model and for independent samples of increasing size $M$. The horizontal dotted line denotes the analytically computed value $\sigma^{(1)}_{3/2} \approx 0.820$. In the case $\alpha=1$ we have $\sigma_1^{(1)}=1/2$. 
    Note that the figure was truncated with a maximum of $5$ for the ordinate, while the standard deviation showed incursions up to $100$.}
    \label{fig:evolsigmas}
\end{figure}

We propose to use the model (that allows for very large samples) in the simple case $n=1$ to evaluate the reliability of this quantity: 
figure \ref{fig:evolsigmas} shows the evolution of the standard deviation $\sigma_2^{(1)}$ for increasing sample size $M$. 
For each sample we draw $M$ times of flight, so that the samples are totally independent. 
As expected, the standard deviation does not converge but continues to increase with the sample size $M$, so that there is no limit value for this quantity, even if the divergence is very slow. 
What is more surprising is the fact that the signal is incredibly noisy: indeed, while we show only data in the reduced vertical range $[0:5]$, values of up to $100$ are present. 
Finally, although the fact that the different samples are independent may explain part of the randomness of the curve, we could expect at least the noise to decay when $M$ increases. 
This is obviously not the case, which means that the standard deviation cannot even be used to compare two different laminar mixers using the same sample size.  
This point has to be stressed since, because of turbulent flows where distributions are close to Gaussian, nearly all RTD studies in fluid mechanics use this parameter (and sometimes higher moments) \cite{bib:ham_etal2004,bib:trachsel2005,bib:adeosun2009,bib:bovskovic_etal2011,bib:rodrigues2021}. 
The difficulty lies indeed in the fact that a logarithmic divergence is extremely difficult to detect from a series of points.  
For an experiment also, the algebraic decay is impossible to monitor in practice, so that the tail --responsible for the logarithmic divergence-- will not be fully taken into account, hiding the problem. 

Since the moment of order 2, related to the standard deviation, is mathematically ill-posed, we propose to use a centered absolute moment of order $\alpha$ defined as 
\begin{equation}
    \sigma_\alpha^{(n)}=\left(\int_{t_\mathrm{min}^{(n)}}^\infty f_n(t)\,\left|\frac{t}{t_m^{(n)}}-1\right|^{\alpha} dt\right)^{1/\alpha} \,,
    \label{eq:sigma_alpha_continuous}
\end{equation}
where $\alpha$ is strictly less than $2$ and can be fractional; 
fractional moments are indeed frequently used in physics for evaluation of heavy-tailed distributions \cite{bib:srokowski2009}. 
In practice, it can also be calculated from a finite series of $M$ values of time of flight, as done for the standard deviation. 
We obtain:
\begin{equation}
\sigma_{\alpha}^{(n)}=\left(\frac{1}{M}\sum_{j=1}^M \left|\frac{t_j}{t_m^{(n)}}-1\right|^\alpha\right)^{1/\alpha}\,.    
\label{eq:sigma_alpha_tof}
\end{equation}
Here again, the weighting is already contained in the Lagrangian nature of the time of flight. Evaluating this quantity from points uniformly distributed at inlet is described later (see equation \ref{eq:mean_absolute}). 

\subsection{Choice of $\alpha$}

In our case, taking $\alpha=1.99$ would do fine in theory, since the integral would converge. 
However, as seen in figure \ref{fig:evolsigmas}, the signal is very noisy for $\alpha=2$, and we expect the chosen quantity to converge reasonably rapidly with increasing $M$. 
We therefore propose to test two different values of $\alpha$, namely $\alpha=3/2$ and $\alpha=1$. 
The moment of order $1$ \cite{bib:khair2017} is more specifically named ``mean absolute deviation'' in statistics.
As for the usual standard deviation, we wish to evaluate the reliability of these quantities using one element of the model ($n=1$). We denote $\sigma_\alpha\equiv\sigma_\alpha^{(1)}$: we will check that the series in Eq. \eqref{eq:sigma_alpha_tof} actually converge when increasing the size $M$ of the sample, and compare how fast they converge toward the limit $\sigma_\alpha$ for the two values of $\alpha$. 
We thus need an analytical expression of $\sigma_\alpha$ from the model flow, calculated from equation \eqref{eq:sigma_alpha_continuous}. 

The case $\alpha=1$ is straightforward and leads to $\sigma_1=1/2$ for the model flow.
Matsui \& Pawlas calculated existing fractional moments of Pareto functions using Laplace transforms \cite{bib:matsui_pawlas2016}; 
the results are expressed in terms of the beta function and the Gauss hypergeometric function.
We give in appendix \ref{app:sigma_3/2_analyt_model} a classic analytical calculation: we obtain $\sigma_{3/2}\approx0.820$ for the model flow, and we expect to find the same value numerically. 

In figure \ref{fig:evolsigmas} we show the evolution of these quantities as a function of the sample size $M$, using the same set of data already used for the standard deviation  $\sigma_2^{(1)}$. 
Whilst both moments converge toward the desired limits, the convergence is far more rapid in the case $\alpha=1$. 
The signal is also much less noisy for the mean absolute deviation, obviously much less sensitive to the presence of very large residence times in the sample. 
Note finally that $\sigma_1$ is reasonably converged for a quite low sample size ($M\ge10^3$--$10^4$).

\subsection{Influence of molecular diffusion}

Since the reason for the divergence of the standard deviation $\sigma_2$ is linked to the existence of arbitrary long residence times, we could wonder whether this phenomenon would be effectively observed when molecular diffusion is taken into account. 
Indeed, molecular diffusion would allow the fluid particle to change streamline, preventing very long residence times from being observed. 
In numerical simulations also, even without diffusion, the calculations would be stopped in the case of too large residence times. 
This cut-off could enable the convergence of the standard deviation, and render this parameter acceptable for calculating the width of distributions. 
In order to evaluate how molecular diffusion would modify the preceding result, we proceed as follows: 
as for figure \ref{fig:evolsigmas}, we consider one element of the Poiseuille model flow, with length  $L=D$, where $D$ is the diameter of the entrance section.  
We define the P\'eclet number of the flow as $Pe=v_m D/D_s$, where $D_s$ is the molecular diffusion of the species considered. 
The displacement of a given diffusing species obeys to
\begin{equation}
    \frac{d\mathbf{x}}{dt}=\mathbf{v}(x,y,z,t)+\mathbf{\zeta} (t)\, , 
    \label{eq:displacement_diffusion}
\end{equation}
where $\zeta(t)$ is a Gaussian decorrelated process such that $\langle \zeta_i(t)\zeta_j(t')\rangle=2 D_s \delta_{ij}\,\delta(t-t')$ 
\cite{bib:Aref_Jones_1989}. 
For the model $\mathbf{v}$ is simply given by equation \eqref{eq:velocity_Poiseuille_circular}. 
As done in figure \ref{fig:evolsigmas}, for an abscissa $M$ we generate $M$ random initial locations with a parabolic probability. 
For those $M$ initial points we solve equation \eqref{eq:displacement_diffusion} between $x=0$ and $x=L$ for different realistic finite P\'eclet numbers ($Pe=10^6$, $10^7$ and $10^8$), and also in the case without diffusion ($Pe=+\infty$); for each case we plot the standard deviation $\sigma_2$ and the mean absolute deviation $\sigma_1$ of the resulting RTD.
\begin{figure}
    \includegraphics[]{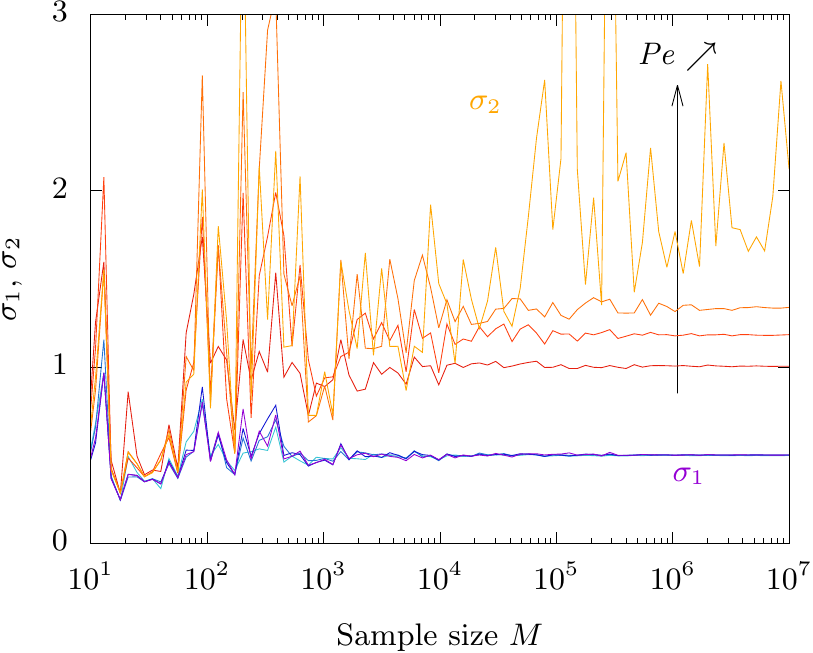}
    \caption{Evolution of the standard deviation $\sigma_2$ (from red to yellow) and mean absolute deviation $\sigma_{1}$ (from blue to purple) for diffusing species in one element of the Poiseuille model flow, and independent samples of increasing size $M$. The P\'eclet numbers are $\Pe = 10^6, \, 10^7, \, 10^8, \, +\infty$. The number of points is here much less than in figure \ref{fig:evolsigmas}, typically 15 samples in a decade compared to 150 in figure \ref{fig:evolsigmas}. Because molecular diffusion should not play a significant role in a single element of mixer, the different plateaus obtained for $\sigma_2$ are clearly artificial, and show again that the standard deviation is ill-posed here. In contrast, the mean absolute deviation $\sigma_1$ is nearly insensitive to molecular diffusion, proving that $\sigma_1$ is a robust measure of the width of RTD for a given mixer. 
     }
    \label{fig:evol_sigma_diff}
\end{figure}
As is visible in figure \ref{fig:evol_sigma_diff}, $\sigma_2$ converges for finite P\'eclet number. 
However, the convergence is slow; even more important, the value of the plateau depends  significantly on the P\'eclet number. 
Although we could expect a small dependence for a long mixer, those large differences for a single piece of mixer at high P\'eclet numbers are not physical, which shows that the converged value obtained for $\sigma_2$ is artificial. 
In the case of $\sigma_1$, the curves  merge for quite small samples, and, as expected in that situation, converge toward the theoretical value $\sigma_1=1/2$, whatever the P\'eclet number. 
This clearly shows that, unlike the standard deviation, the mean absolute deviation is a robust measure of the width of the distributions.

\subsection{Application to the different mixers}
Because the number of numerical data points used for the different mixers is in between $17\,000$ and $34\,000$, from the analysis above we have enough data to calculate reasonably accurately the
mean absolute deviation, 
\begin{equation}
    \sigma_1^{(n)}=\frac{1}{M}\sum_{j=1}^M \left|\frac{t_j}{t_m^{(n)}}-1\right|\,.
    \label{eq:sigma_1^n_non_weighted}
\end{equation}

\begin{figure}
    \includegraphics[]{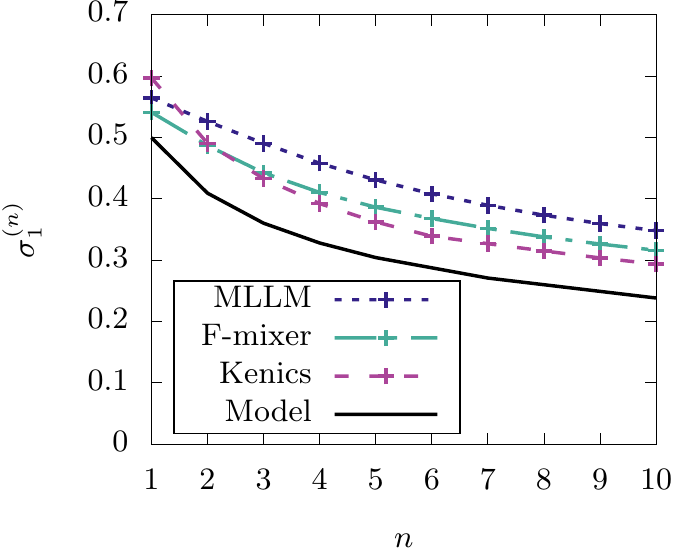}
    \caption{Evolution of the mean absolute deviation $\sigma_{1}^{(n)}$ with the number of elements $n= $ 1, 2, 3, 4, 5, 7, 10, for the three mixers and the model.  }
    \label{fig:evol_mom1}
\end{figure}

Figure \ref{fig:evol_mom1} shows $\sigma_1^{(n)}$ as a function of the number of elements for the three mixers and the model. 
In appendix \ref{app:sigma_3/2}, we show the same evolution for $\sigma_{3/2}^{(n)}$; we can check that the hierarchy between the  different mixers is the same for the two different values of $\alpha$, which definitely reinforces the choice $\alpha=1$.

Without surprise, the totally uncorrelated model is the most efficient. 
As expected also, the reduced moment of the MLLM, which is the most correlated mixer, decreases less rapidly than the others; 
the Kenics$^{\hbox{\textregistered}}$ is the best of the three mixers from the RTD point of view. 

Although $\sigma_1^{(n)}$ is a decreasing function of $n$ for all cases considered, there is no obvious analytical fit for the decay even in the case of the decorrelated model.
The decrease is the most rapid at the beginning, for small values of $n$: 
the width of the distributions (measured with  $\sigma_1^{(n)}$) has decreased by 25\% (for the MLLM) to 40\% (for the model) after $n=5$ elements, but the decrease is only 40 to 52\% for $n=10$. 
Hence from the RTD point of view there is no interest in adding many elements in a row, provided that a good mixing is reached after a few number of elements.

\section{Influence of the cross-section geometry}
\label{sec:discussion}
In this section we would like to understand the reason for the differences in values of  $\sigma_1^{(n)}$. 
Since large times of flight are linked to the presence of walls, one could wonder whether the shape of the mixer is of importance. 
As noted by Mortensen \textit{et al.} \cite{bib:mortensen_etal2005}, a shape can be characterized by a perimeter ${\cal P}$ and an area ${\cal A}$, that can be combined in a dimensionless compactness number ${\cal C}$, defined as 
\begin{equation}
    {\cal C}=\frac{{\cal P}^2}{{\cal A}}\,.
\end{equation}
This quantity is not easy to measure for the mixers considered here. 
We thus propose to consider model flows as the one proposed in section \ref{sec:residence_time_model}.  
We formerly took the case of a circular cross section, which allowed for analytical exact results easily comparable to numerical simulations. 
But it is relatively simple to investigate  different compactness by varying the shape of the cross-section (ellipse, square, or rectangle rather than a circle), as done in Mortensen \textit{et al.} \cite{bib:mortensen_etal2005}. 
Because $\sigma_1^{(n)}$ decays roughly similarly with $n$ for all mixers (Fig. \ref{fig:evol_mom1}), we focus on the value $n=1$. 

In the following we keep the area ${\cal A}$, length $L$ and the flow rate ${\cal Q}$ constant, so that all different shapes correspond to the same mean time $t_m$.

\subsection{Ellipse}
There is no exact expression for the perimeter of an ellipse; 
however it can be approximated using Ramanujan's second formula \cite{bib:ramanujan1962}:
\begin{equation}
\!\!\!{\cal P}\approx\pi(a+b)\left(1+\frac{3\lambda^2}{10+\sqrt{4-3\lambda^2}}\right) \hbox{ with } \lambda=\frac{a-b}{a+b}
\label{eq:perimeter_ellipse}
\end{equation}
where $a$ and $b$ are the large and small semi-axes respectively. 
This expression is very accurate, even for very elongated ellipses \cite{bib:Villarino2005}. 
The parameter $\lambda$ varies from $0$ (circle) to $1$ (very elongated ellipses).
If ${\cal A}=\pi a b$ is kept constant, then $a+b=a+{\cal A}/(\pi a)$ is minimum for the circle; 
the bracketed expression in equation \ref{eq:perimeter_ellipse} is also a growing function of $a$, so that the perimeter is always increasing with $a$.  
The area ${\cal A}$ being kept constant, the compactness number ${\cal C}$ also increases with $a$. 

However, as shown in appendix \ref{sec:ellipse_RTD}, the probability density of time duration for a pipe of length $L$ is identical for a circular or elliptic cross-section, whatever $\lambda$. 
This implies that all moments derived (including  $\sigma_1^{(1)}$) are identical. 
In this case the compactness ${\cal C}$ plays no role on the distribution of duration times. 
Nevertheless, $\sigma_1^{(1)}$ may depend on the geometry, number of angles, \textit{etc.}

\subsection{Square and rectangles}

Let us consider the Hagen-Poiseuille flow with rectangular cross-section. 
The rectangle has a width $a$, a height $b$, and is characterized by its area ${\cal A}=a\times b$ and aspect ratio $\beta = b/a$. 
For this configuration, Spiga and Morino \cite{bib:spiga2006} proposed the following expression for the velocity field:
\begin{align}
&v(y,z) = \frac{16 a^2 b^2 G}{\mu \pi^4} \times \nonumber \\ &\sum_{n \, \rm{odd}}^{\infty} \sum_{m \, \rm{odd}}^{\infty} \frac{\sin[n \pi (y/a - 1/2)]\, \sin[ m \pi(z/b - 1/2) ]}{n m\, (b^2 n^2 + a^2 m^2)}
\end{align}
for $-a/2 \leqslant y \leqslant a/2$ and $-b/2 \leqslant z \leqslant b/2$, with $G$ the imposed pressure gradient and $\mu$ the dynamic viscosity of the fluid. 
The mean velocity $v_m$ is therefore:
\begin{equation}
v_{m} = \frac{64 a^2 b^2 G}{\mu \pi^6} \sum_{n \, \rm{odd}}^{\infty} \sum_{m \, \rm{odd}}^{\infty} \frac{1}{n^2 m^2 (b^2 n^2 + a^2 m^2)}
\end{equation}
In practice, this series converges rather rapidly, and we checked that truncating the sums such that $0\le n,m\le1000$ was enough for our calculation.
The aspect ratio is varied from $\beta = 1$ (square cross-section) to $\beta = 1000$ (very elongated rectangle), the limit $\beta \rightarrow +  \infty$ being the plane Poiseuille flow. 
Finally all times are made non dimensional using the mean time $t_m=L/v_m$, where $L$ is the length of the pipe section.

Due to the complexity of the expression of the velocity field, the inversion method is of no use in this situation. We can nonetheless compute the mean absolute deviation corresponding to this velocity field by taking points uniformly distributed in the rectangle and weighting the values using the velocity, which modifies expression (\ref{eq:sigma_1^n_non_weighted}) as follows:
\begin{equation}\label{eq:mean_absolute}
\sigma_{1}^{(1)}=\frac{1}{M}\sum_{j=1}^M \frac{v_j}{v_m}\left|\frac{t_j}{t_m^{(1)}}-1\right|   \, .
\end{equation}
This approach was tested on the circular Poiseuille flow, by taking points uniformly distributed on the disk and using the expression (\ref{eq:mean_absolute}), and the same value of  $0.5$  was obtained for $\sigma_{1}^{(1)}$, confirming the validity of the method.

\begin{figure}
    \includegraphics[]{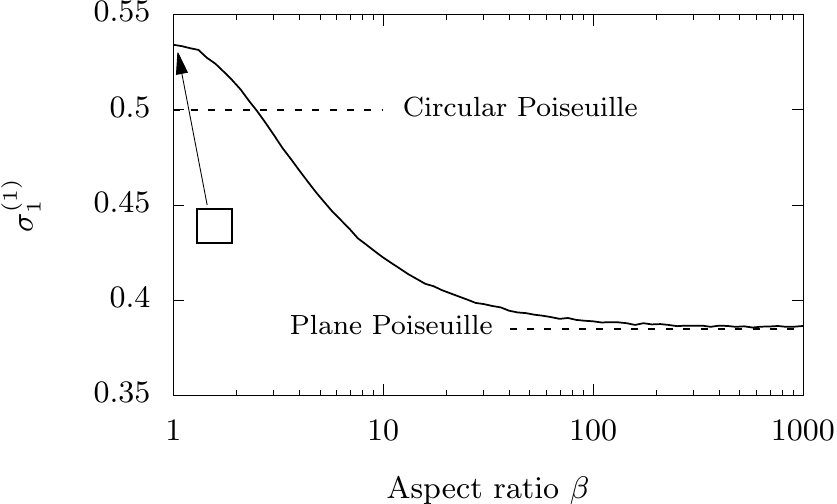}
    \caption{Evolution of the mean absolute deviation for the Hagen-Poiseuille flow in rectangular ducts of varying aspect ratio $\beta$. Dashed lines indicate values of $\sigma^{(1)}_{1}$ for the circular and plane Poiseuille flow configurations respectively $\sigma^{(1)}_{1}=1/2$ and $\sigma^{(1)}_{1}=2/(3\sqrt{3}) \simeq 0.385$. As $\beta$ increases, the configuration tends to that of the plane Poiseuille flow.
    \label{fig:evol_sigma_rect}}
\end{figure}

Figure \ref{fig:evol_sigma_rect} represents the evolution of $\sigma_{1}^{(1)}$ with the aspect ratio $\beta$.
For each value of $\beta$, 3 samples of $100 \, 000$ points were computed, leading to slightly different values of  $\sigma_{1}^{(1)}$ due to the randomness of the process. 
However, because of the rapid convergence of $\sigma_1$ with the sample size, the 3 values are very close to each other, with a typical variation of order 0.5\%; the quantity plotted in figure \ref{fig:evol_sigma_rect} is the mean for those three sets.

We observe that the mean absolute deviation decreases as the aspect ratio increases, converging to the value corresponding to the plane Poiseuille flow.

Note finally that in microfluidics, most microchannels have rectangular cross-section; the case of the circular cross-section, better than the square from the residence time point of view, is very close to the 3-1 rectangle, a geometry quite common in microfluidics.

\section{Summary and conclusion}
In this article we have studied the statistics of residence time distributions for $n$ elements of an in-line mixer, using numerical data for three mixers and a model flow. 
We have shown that those types of mixers are not perfect from the RTD point of view, and that the $t^{-3}$ tail found for one element of mixer persists when increasing the number of elements. 
This algebraic decay, signature of a ``heavy-tailed'' distribution, has an important consequence in practice: 
the second order moment of the distributions --and therefore higher moments-- do not exist, so that the standard deviation cannot be used to characterize the width of the histogram. 

Therefore we proposed to use the first order absolute moment, also called mean absolute deviation, given by equation (\ref{eq:sigma_1^n_non_weighted}): 
this moment exists and converges with increasing sample size in numerical simulations, and should also be used in experiments, where the tail is difficult to obtain in practice. 

The mean absolute deviation is then used to compare the different mixers, and how the typical width of the distribution decreases with $n$. 
It is also applied to discriminate between different shapes of cross-section. 
We show that this parameter is higher for a square than for a circle, but also that a rectangular cross-section, very common in microfluidics, is a better mixer than a square from the RTD point of view. 

One could wonder how the results for a mixer consisting of $n$ elements would be affected by molecular diffusion. 
In fact, molecular diffusion has negligible effects as long as the Batchelor scale is not reached \cite{bib:raynal_gence1997}. 
Since such in-line mixers reproduce the baker's map, the width of a given heterogeneity at the exit of the $n$-th element is typically $\ell_n\sim w/2^n$, where $w$ is the width of the cross-section. 
Such an heterogeneity is mixed on a time-scale $\tau_n\sim \ell_n^2/D_s$.  
where $D$ is the molecular diffusion of the species to be mixed. 
Thus the scalar is mixed at the exit of the $(n+1)$-th element if $\tau_n$ is of the order of the mean travel time in one element $t_m=L/v_m$. 
When equating $\tau_n\sim t_m$, we obtain
\begin{equation}
    2n\ln 2\sim \ln\left(\frac{v_m w}{D_s}\,\frac{w}{L}\right)\,,
\end{equation}
where we recognize the P\'eclet number $Pe=v_m w/D_s$. 
In an in-line mixer, the length $L$ is a few times the width $w$ (see figure \ref{fig:real_mixer_drawing}), whilst the P\'eclet number is typically of the order of $10^6$, so that $\ln(w/L)$ can be neglected in front of $\ln Pe$. 
We finally obtain
\begin{equation}
    n\approx\frac{\ln Pe}{2\ln 2}\,.
\end{equation}
For $Pe=10^6$, we obtain $n\approx10$, so that in that case the effects of diffusion are negligible until the outlet of the mixer. 
In any case, even if the diffusion effects became important in the very last elements, this would not significantly change the statistics of the residence time on the whole mixer, so that our results should apply even if  diffusion is taken into account. 

\section*{Acknowledgments} 
The support from the PMCS2I of \'Ecole centrale de Lyon for the numerical calculations is gratefully acknowledged. 
We also thank an anonymous referee for interesting suggestions and challenging comments.

\appendix

\section{Generation of a random variable with parabolic density} 
\label{app:inversion_method}

This basic technique is described in \cite{bib:devroye1986}. 
The goal is to derive a two-dimensional probability density function (pdf) that is proportional to the velocity field, here in the case of circular cross-section:
\begin{equation}
    f(M) \propto v(M) = v(r)
\end{equation}
By cylindrical symmetry, this is readily reduced to finding a one-dimensional pdf of the variable $r$, that has however to be proportional to the velocity field and the perimeter corresponding to the position considered:
\begin{equation}
    f(r) \propto 2 \pi r v(r) = 2 \pi r \times 2 v_m \left( 1 - (r/R)^2 \right)
\end{equation}
Since the integral of $f$ over $[0,R]$ has to be $1$, we easily obtain:
\begin{equation}
    f(r) = \frac{4r}{R^2} \left( 1 - \frac{r^2}{R^2} \right) 
\end{equation}
We then compute the corresponding cumulative density function (cdf) $F$, primitive of $f$:
\begin{equation}
    F(r) = \frac{2r^2}{R^2} \left( 1 - \frac{r^2}{2R^2} \right) 
\end{equation}
Finally, the inverse function of $F$ is expressed as:
\begin{equation}
    F^{-1}(p) = R \sqrt{1 - \sqrt{1-p}} \, , \quad \forall p \in [0,1[
\end{equation}
From here, the inversion method consists in generating a sample $(p_i)_{1 \leqslant i \leqslant M}$ of reals uniformly distributed between 0 and 1; 
in practice, we use a pseudo-random numbers generator (PRNG) to produce the uniform distribution, here the \texttt{xoshiro256**} PRNG of the \texttt{gfortran} compiler.
We then apply $F^{-1}$ to the sample produced. 
The result is a new sample of radii $(r_i)_{1 \leqslant i \leqslant M} = (F^{-1}(p_i))_{1 \leqslant i \leqslant M}$ which follows the distribution law described by $f$. 

\section{Tail of RTD for \texorpdfstring{$n$\ }\ identical elements}
\label{app:tail}
Suppose that the RTD of 1 element of a mixer possesses a $t^{-3}$ tail. 
Then, if the elements are decorrelated from the residence time point of view, the tail of the distribution of $n$ elements also has a $t^{-3}$ tail.  

\begin{proof}
We will proceed by recurrence.
We denote by $f_m(t)$ the pdf associated to the crossing time for $m$ sections. 
We suppose that for all $m\le (n-1)$, we have 
\begin{equation}
f_m(t)=\frac{g_m(t)}{(t+t_\epsilon)^{-3}}
\end{equation}
where $t_\epsilon$ is an arbitrary positive time, and $g_m(t)$ a smooth function such as
\begin{equation}
g_m(t<m\times t_{min})=0 \ \hbox{ and } \lim_{t\rightarrow\infty} g_m(t)=C_m\not=0\,.
\label{eq_assertation_gm}
\end{equation}
The assertion (\ref{eq_assertation_gm}) is true for $n=1$; we suppose that it is also true for $n-1$ and prove that it is true for $n$. 
Providing that the events are sufficiently decorrelated, the pdf for $n$ elements is the convolution product of $f_1$ with $f_{n-1}$:
\begin{eqnarray}
f_n(t)&=&\int_{-\infty}^{+\infty}f_1(t)\,f_{n-1}(t_n-t)\,dt\nonumber\\
&=&\int_{t_{min}}^{t_n-(n-1)t_{min}}\!\!\!\!\frac{g_1(t)}{(t+t_\epsilon)^3}\,\frac{g_{n-1}(t_n-t)}{(t_n+t_\epsilon-t)^3}\,dt
\end{eqnarray}
We make the change of variable $x=t/t_n$, so that $dt=t_n\,dx$:
\begin{eqnarray}
 f_n(t)\!\!\!\!&=&\!\!\!\!\!\!\int_{t_{min}/t_n}^{1-(n-1)t_{min}/t_n}\!\!\!\!\!\!\frac{g_1(t_nx)}{(t_nx+t_\epsilon)^3}\,\frac{g_{n-1}(t_n(1-x))}{(t_\epsilon+t_n(1-x))^3}\,t_n\,dx\nonumber\\
&\underset{t_n\rightarrow\infty}{\sim}&\int_0^1t_n^{1-3-3}\frac{g_1(t_nx)}{(x+t_\epsilon/t_n)^3}\,\frac{g_{n-1}(t_n(1-x))}{(t_\epsilon/t_n+1-x)^3}\,dx\nonumber\\
&\underset{t_n\rightarrow\infty}{\sim}&t_n^{-5}\int_0^1\frac{g_1(t_nx)}{(x+t_\epsilon/t_n)^3}\,\frac{g_{n-1}(t_n(1-x))}{(t_\epsilon/t_n+1-x)^3}\,dx \label{eq:equivalent_p_n_intermediaire}
\end{eqnarray}
because of the presence of the constant $t_\epsilon$, the function to integrate remains smooth on $[0;1]$. 
Let us focus on equation (\ref{eq:equivalent_p_n_intermediaire}): 
when $t_n\rightarrow\infty$, we have $t_\epsilon/t_n\rightarrow0$, and we have 2 important contributions, one at $x=0$ and the other at $x=1$. 
We thus neglect other contributions: 
in the vicinity of $x=0$, the function to integrate is equivalent to $A_0\, (x+t_\epsilon/t_n)^{-3}$, and in the vicinity of $x=1$, is equivalent to $A_1 (t_\epsilon/t_n+1-x)^{-3}$, where the fonctions that do not tend to infinity have been approximated by constants. 
We obtain:
\begin{eqnarray}
f_n(t)&\underset{t_n\rightarrow\infty}{\sim}& t_n^{-5}\left[ \frac{A_0}{2(t_\epsilon/t_n)^2}+\frac{A_1}{2(t_\epsilon/t_n)^2}\right]\\
&\underset{t_n\rightarrow\infty}{\sim}& C_n\, t_n^{-3}
\end{eqnarray}
We have shown that $f_n(t)$ also has a $t^{-3}$ tail and, by recurrence, the property is true for all $n$. 
\end{proof}

\section{Calculation of the reduced moment \texorpdfstring{$\sigma_{3/2}^{(1)}$\ }\ for the model for 1 element of a cylindrical pipe}
\label{app:sigma_3/2_analyt_model}

The reduced moment $\sigma_{3/2}^{(1)}$ for one element of the model writes:
\begin{equation}
\sigma_{3/2}^{(1)}=\left(\frac{\sqrt{t_m}}{2}\int_{t_m/2}^\infty \frac{|t-t_m|^{3/2}}{t^3}\; dt\right)^{2/3}
\end{equation}
Because of the absolute value, the integral is divided, one integral for $t\le t_m$ (denoted by $I_1$) and the other for $t\ge t_m$ (denoted by $I_2$), such that
\begin{equation}
\sigma_{3/2}^{(1)}=\left(I_1+I_2\right)^{2/3}\,.  
\end{equation}

\textbf{Calculation of $I_1$ $(t\le t_m)$:}\\
we set $u^2=(1-t/t_m)$. $I_1$ satisfies
\begin{equation}
I_1=\int_0^{1/\sqrt{2}}\frac{u^4}{(1-u^2)^3}\, du
\end{equation}
We use formula 2.147(4) page 77 from Gradshteyn \& Ryzhik \cite{bib:gradshteyn2007}:
\begin{eqnarray}
&\hspace{-1.5cm}\displaystyle\int\frac{x^m\, dx}{(1-x^2)^n}=&\frac{1}{2n-2}\frac{x^{m-1}}{(1-x^2)^{n-1}}\nonumber\\
&&\hspace{0.5cm}-\frac{m-1}{2n-2}\int\frac{x^{m-2}\,dx}{(1-x^2)^{n-1}}
\end{eqnarray}
first with $m=4$ and $n=3$, next with $m=n=2$ and obtain:
\begin{equation}
I_1=-\frac{1}{4\sqrt{2}}+\frac{3}{8}\ln(\sqrt{2}+1)
\end{equation}
\textbf{Calculation of $I_2$ $(t\ge t_m)$:}\\
we set $u^2=(t/t_m-1)$ and obtain:
\begin{equation}
I_2=\int_0^{\infty}\frac{u^4}{(1+u^2)^3}\, du
\end{equation}
We next use formula 3.241(4) page 322 from Gradshteyn \& Ryzhik \cite{bib:gradshteyn2007}:
\begin{eqnarray}
&\hspace{-1.5cm}\int_0^\infty \frac{x^{\mu-1}\,dx}{(p+qx^\nu)^{n+1}}=\frac{1}{\nu p^{n+1}}\left(\frac{p}{q}\right)^{\mu/\nu}\frac{\Gamma(\mu/\nu)\,\Gamma(1+n-\mu/\nu)}{\Gamma(1+n)}
\end{eqnarray}
with $\mu=5$, $\nu=2$, $p=q=1$ and $n=2$:
\begin{eqnarray}
I_2&=&\frac{1}{2}\frac{\Gamma(5/2)\,\Gamma(1/2)}{\Gamma(3)}\\
&=&\frac{3\pi}{16}
\end{eqnarray}

We finally obtain:
\begin{eqnarray}
\sigma_{3/2}^{(1)}&=&\left(-\frac{1}{4\sqrt{2}}+\frac{3}{8}\ln(\sqrt{2}+1)+\frac{3\pi}{16}\right)^{2/3}\nonumber\\
&\approx&0.820\,.
\end{eqnarray}
Ramsay \cite{bib:ramsay2006} calculated fractional moments of this type of distribution using Laplace transforms, and gave the result in the form of an infinite series. 
We checked that the series indeed converged toward the same value. 

\section{Evolution of the reduced moment  \texorpdfstring{$\sigma_{3/2}^{(1)}$\ }\  with the number of elements}
\label{app:sigma_3/2}
\begin{figure}[h!]
    \includegraphics[]{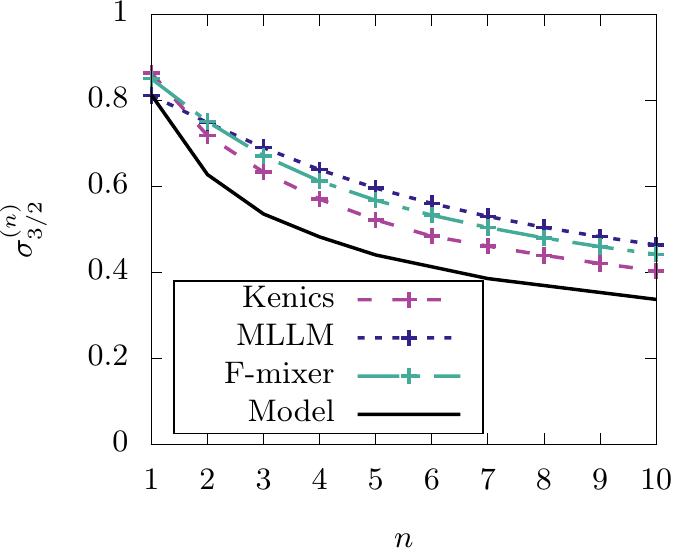}
    \caption{Evolution of $\sigma_{3/2}^{(n)}$ with the number of elements $n= $ 1, 2, 3, 4, 5, 7, 10, for the three real mixers and the model.}
    \label{fig:evol_mom1dot5}
\end{figure}
The evolution of the reduced moment of order $3/2$ with the number $n$ of elements is shown in figure \ref{fig:evol_mom1dot5}.
When compared to figure \ref{fig:evol_mom1}, the hierarchy between the different mixers is preserved; the decay with $n$ is also similar.

\section{From circular to elliptic cross-section: calculation of RTD for 1 element of model mixer}
\label{sec:ellipse_RTD}
The velocity field for an ellipse of semi-axes $a$ and $b$ writes:
\begin{equation}
v_x(y,z)=2v_m\left(1-\frac{y^2}{a^2}-\frac{z^2}{b^2}\right)\,,
\label{eq:vitesse_ellipse}
\end{equation}
where $v_m$ denotes the the mean velocity. 
we denote by $g(t)$ the density probability to have a time of flight of duration $t$ for an element of size $L$, with
\begin{equation}
t=L/v_x(y,z)\,.
\label{eq:t_L_V}
\end{equation}

\begin{figure}[h!]
\includegraphics[]{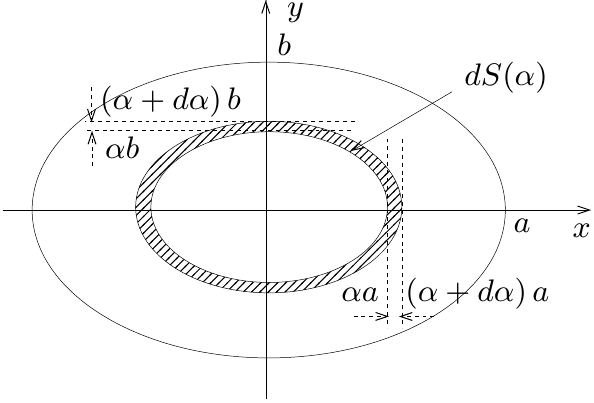}
\caption{Elliptic case: the hatched area represents the surface between the two ellipses corresponding to values $\alpha$ and $\alpha+d\alpha$ in equation \ref{eq:vitesse_ellipse_alpha}.}
\label{fig:ellipses}
\end{figure}

Let us consider the points that verify
\begin{equation}
\frac{y^2}{a^2}+\frac{z^2}{b^2}=\alpha^2, \quad 0\le\alpha\le 1\,.
\label{eq:ellipse_alpha}
\end{equation}
They describe an ellipse of axes $\alpha a$ and $\alpha b$. 
From equations \ref{eq:vitesse_ellipse}, \ref{eq:t_L_V} and \ref{eq:ellipse_alpha} we obtain
\begin{equation}
v_x(\alpha)=2v_m\,(1-\alpha^2)=L/t\,,
\label{eq:vitesse_ellipse_alpha}
\end{equation}
that differentiates into
\begin{equation}
4\alpha d\alpha \,v_m=L\, dt/t^2\,.
\label{eq:vitesse_ellipse_alpha_differencie}
\end{equation}
Because the density probability is proportional to the velocity, we now write that the probability that $t$ is in between $t$ and $t+dt$ is the same as that for $v_x$ to be in between $v_x(\alpha)$ and $v_x(\alpha+d\alpha)$:
\begin{equation}
g(t)\,dt= \frac{v_x(\alpha)}{v_m}\,\frac{dS(\alpha)}{\pi\,ab}\,,
\label{eq:g_t_v_alphadS_alpha}
\end{equation}
with $dS(\alpha)$ the surface difference between ellipses corresponding to $\alpha+d\alpha$ and $\alpha$, see figure \ref{fig:ellipses}. 
We thus have:
\begin{eqnarray}
dS(\alpha)&=&\pi a b\bigl[(\alpha+d\alpha)^2-\alpha^2\bigr]\nonumber\\
&\approx& 2 \pi a b \, \alpha\, d\alpha\label{eq:ellipse_surf_dalpha}
\end{eqnarray}
By combining equations (\ref{eq:vitesse_ellipse_alpha}), (\ref{eq:vitesse_ellipse_alpha_differencie}), (\ref{eq:g_t_v_alphadS_alpha}) et (\ref{eq:ellipse_surf_dalpha}), we obtain
\begin{eqnarray}
g(t)&=&\frac{L}{v_m t}\,2 \,\alpha\,  \frac{d\alpha}{dt}\\
&=& \frac{L^2}{2v_m^2,t^3}\,. 
\end{eqnarray}
The mean time $t_m$ verifies
\begin{eqnarray}
t_m&=&\frac{1}{\pi a b}\int_{\alpha=0}^{\alpha=1}\frac{v_x(\alpha)}{v_m}\times\frac{L}{v_x(\alpha)}\,dS(\alpha)\\
&=&\frac{L}{v_m}\,.
\end{eqnarray}
We finally obtain
\begin{equation}
g(t)=\frac{t^2_m}{2\,t^3}\,,
\end{equation}
that is, the same expression as for the circle equation \ref{eq:f1_Poiseuille_circular}. 

\vfill


\bibliographystyle{unsrt}

\bibliography{biblio}

\end{document}